\documentclass{iopjournal}
\usepackage{amsmath}
\usepackage{url}
\usepackage{xcolor}
\begin{document}

\articletype{Article type} 

\title{Two-dimensional beam compression for sub-femtosecond electron beam generation}

\author{Weihang Liu$^{1,2,\dagger}$, Shimin Jiang$^{1,2,\dagger}$, Xiao Li$^{1,2,*}$, Xingguang Liu$^{1,2,*}$, Yi Jiao$^{1,*}$ and Sheng Wang$^{1,2,*}$}

\affil{$^1$Institute of High Energy Physics, Chinese Academy of Sciences, Beijing 100049, China}

\affil{$^2$Spallation Neutron Source Science Center, Dongguan 523803, China}

\affil{$^\dagger$These authors contributed equally to this work.}

\affil{$^*$Authors to whom any correspondence should be addressed.}

\email{lixiao@ihep.ac.cn, liuxg@ihep.ac.cn, jiaoyi@ihep.ac.cn, wangs@ihep.ac.cn}


\begin{abstract}
Sub-femtosecond electron beams are powerful probes of ultrafast electronic, atomic, and nuclear dynamics, and promising drivers for ultrashort radiation generation from the extreme-ultraviolet to gamma-ray regimes. However, producing such beams at hundred-MeV energies with pC-level charge remains challenging. Here we {develop} a two-dimensional beam-compression scheme based on transverse--longitudinal coupling, in which dispersive beam optics convert the small transverse emittance of modern electron beams into an ultrashort bunch length. Linear analysis and particle tracking show that, after the dominant longitudinal and energy-spread contributions are cancelled, the compressed bunch length is governed primarily by transverse beam quality and collective-effect growth. {We further derive and verify a scaling model that captures collective-effect-induced bunch-length degradation and provides a charge--energy operating map for sub-femtosecond compression.} Start-to-end simulations of a realistic injector-to-compressor beamline produce a 200 MeV, pC-level bunch with a bunch length of 0.45 fs and a peak current of about 3.5 kA. {Jitter studies over a broad range of beam energies show that the output bunch-length distribution narrows and then varies only weakly with increasing energy.} These results suggest a feasible route toward compact, high-energy attosecond electron beam sources and may provide a basis for future sub-femtosecond radiation sources based on undulator emission or inverse Compton scattering.
\end{abstract}

\section{Introduction}
High-quality electron beams are essential tools for a broad range of scientific applications. In accelerator-based light sources, colliders, and ultrafast electron scattering facilities, improvements in beam quality directly translate into enhanced performance, higher brightness, and improved temporal or spatial resolution. Among the many parameters that characterize an electron beam, the bunch length has attracted particular attention~\cite{Zhou2021}. In light-source applications and ultrafast electron diffraction, shorter electron bunches can provide higher temporal resolution and enable the investigation of faster dynamical processes.

Femtosecond electron bunches can now be generated routinely using radio-frequency acceleration and magnetic compression techniques. However, attosecond-long electron pulses had not been available until very recently, and the generation of sub-femtosecond electron beams remains highly challenging~\cite{Morimoto2023Review}. When the target bunch length approaches the sub-femtosecond regime, conventional longitudinal compression methods encounter two major limitations.

The first limitation is associated with beam energy~\cite{PhysRevApplied.15.044039}. For high-energy electron beams, purely longitudinal compression generally requires a very large correlated energy chirp, which significantly increases the cost and complexity of the accelerator system. As a representative example, consider an electron beam with an energy of several hundred MeV and a typical relative energy spread of about $10^{-4}$ from a photocathode injector. To compress such a beam to a bunch length of approximately 1 fs, a C-band accelerating cavity voltage on the order of hundreds of MV would be required. This voltage is already sufficient to increase the beam energy by several times. Assuming a typical C-band accelerating gradient of about 40 MV/m, the required cavity length would be on the order of tens of meters. Such requirements make conventional longitudinal compression an impractical route toward sub-femtosecond bunch generation at high beam energies.

The second limitation is associated with bunch charge~\cite{3t5l-mz61}. Existing non-plasma-based sub-femtosecond electron-pulse generation schemes typically operate at relatively low beam energies, where collective effects, especially space-charge forces, become increasingly important and can strongly degrade the beam quality during compression. Consequently, experimentally achievable sub-femtosecond electron pulses are generally limited to very low bunch charges, typically below the pC level.

The combined constraints on beam energy and bunch charge further limit the scientific impact of existing sub-femtosecond electron-beam generation methods. At present, their applications are mainly confined to ultrafast electron diffraction and related low-charge experiments~\cite{RevModPhys.94.045004}. In contrast, sub-femtosecond electron beams with both higher energy and higher charge could open new opportunities in several areas. For example, they may enable undulator-based sub-femtosecond radiation sources from the extreme-ultraviolet to the x-ray regime, providing a powerful probe of electronic dynamics. They may also be used to generate sub-femtosecond gamma-ray pulses through inverse Compton scattering, with potential applications in nuclear physics and related fields~\cite{PhysRevLett.128.162501,PhysRevLett.128.212502}. Therefore, a method capable of producing high-energy, pC-level, sub-femtosecond electron bunches would be of considerable interest.

{
Transverse--longitudinal coupling (TLC) provides an alternative route to bunch
compression by exploiting the small transverse emittance of an electron beam.
Several classes of TLC-based beam-manipulation schemes have been developed,
including phase-merging enhanced harmonic generation~\cite{PhysRevLett.111.084801}
and angular-dispersion-induced microbunching (ADM)~\cite{Feng2017}.
Although these schemes were originally introduced mainly for harmonic generation
and coherent-radiation applications, they can also operate as bunch compressors.

In particular, Deng \textit{et al.} generalized the TLC analysis and showed that
the final bunch length can be determined by the initial transverse beam size or
angular divergence rather than by the initial energy spread~\cite{Deng2021}.
However, that theoretical work did not provide a specific, realistic beamline
for generating an isolated sub-femtosecond electron bunch. Moreover,
higher-order beam-transport nonlinearities and collective effects were not
systematically investigated. These effects are expected to become particularly
important as the entire electron bunch is compressed toward the
sub-femtosecond regime.

Unlike conventional longitudinal compression, which relies primarily on the
initial energy spread and a large longitudinal chirp, TLC-based compression
uses coupled transverse and longitudinal beam dynamics so that the final bunch
length is governed primarily by the transverse geometric emittance rather than
by the initial energy spread. This feature allows the extremely small transverse
emittance of modern electron beams to be used as an effective resource for
generating ultrashort bunches. Since the compression involves coupled motion
in the horizontal and longitudinal phase spaces, we refer to this class of
methods as two-dimensional beam compression.

More recently, Cheng \textit{et al.} proposed an ADM-based two-dimensional
compression scheme for an MeV ultrafast-electron-diffraction beamline
~\cite{3t5l-mz61}. Their method combines pulse-front-tilted photocathode
illumination, THz-driven energy chirping, a dogleg dispersive section, and
subsequent isochronous transport, and produces a bunch length of
\(810~\mathrm{as}\) for a \(3.18~\mathrm{MeV}\) electron beam. This result
demonstrates the feasibility of ADM-based attosecond compression at MeV
energies.
Its extension to substantially higher beam energies is nevertheless challenging
with present THz technology. Furthermore, space-charge forces and nonlinear effects impose more stringent constraints on
beam manipulation at MeV energies. In that scheme, the final bunch charge is only \(6~\mathrm{fC}\). The simulations also show that nonlinear energy modulation
increases the bunch length from the theoretical estimate of approximately
\(600~\mathrm{as}\) to \(810~\mathrm{as}\). Therefore, the generation of
high-energy, pC-level, sub-femtosecond electron bunches through two-dimensional
compression remains an open problem.
\par
}
\newpage
{
In this work, we present a specific and realistic two-dimensional
beam-compression scheme for generating an isolated sub-femtosecond electron
bunch. The proposed scheme extends previous TLC and ADM concepts to the
high-energy, pC-level regime. It uses a rectangular dipole to introduce
transverse--longitudinal coupling, a C-band RF cavity to provide the linear
energy chirp, a K-band linearizer to compensate the dominant second-order
contribution, and a three-dipole dispersive section (TDS) to complete the
compression.

Compared with the conventional dogleg dispersive section used in ADM, the TDS relaxes the direct constraint
linking the RF chirp to the bending angle of the first dipole, thereby broadening
the feasible parameter space and enhancing the compression performance. In addition to developing the
linear compression theory, we systematically investigate the effects of
higher-order beam-transport nonlinearities, incoherent synchrotron radiation
(ISR), coherent synchrotron radiation (CSR), and space charge (SC). We also derive
a scaling relation that connects the compressed bunch length with the bunch
charge and beam energy.

Start-to-end simulations show that the proposed two-dimensional
beam-compression method can produce a \(200~\mathrm{MeV}\), \(4~\mathrm{pC}\)
electron bunch with a bunch length of \(0.45~\mathrm{fs}\) and a peak current
of approximately \(3.5~\mathrm{kA}\). The present work therefore provides a
practical route for extending two-dimensional beam compression from a general
theoretical concept and a low-energy, fC-level implementation to the
hundred-MeV, pC-level regime.
\par
}

The remainder of this paper is organized as follows. In Sec. II, we present the principle of two-dimensional beam compression, including both the linear theory and the effects of nonlinear terms, and provide a proof-of-principle demonstration. {In Sec. III, we quantify collective-effect-induced bunch-length degradation, develop a scaling model for its dependence on bunch charge and beam energy, and establish the corresponding sub-femtosecond operating map.} In Sec. IV, we present start-to-end simulation results for generating sub-femtosecond electron bunches and evaluate the robustness of the scheme through a jitter analysis. Finally, Sec. V summarizes the main conclusions of this work.

\section{Compression method }
\begin{figure}
 \centering
        \includegraphics[width=1\textwidth]{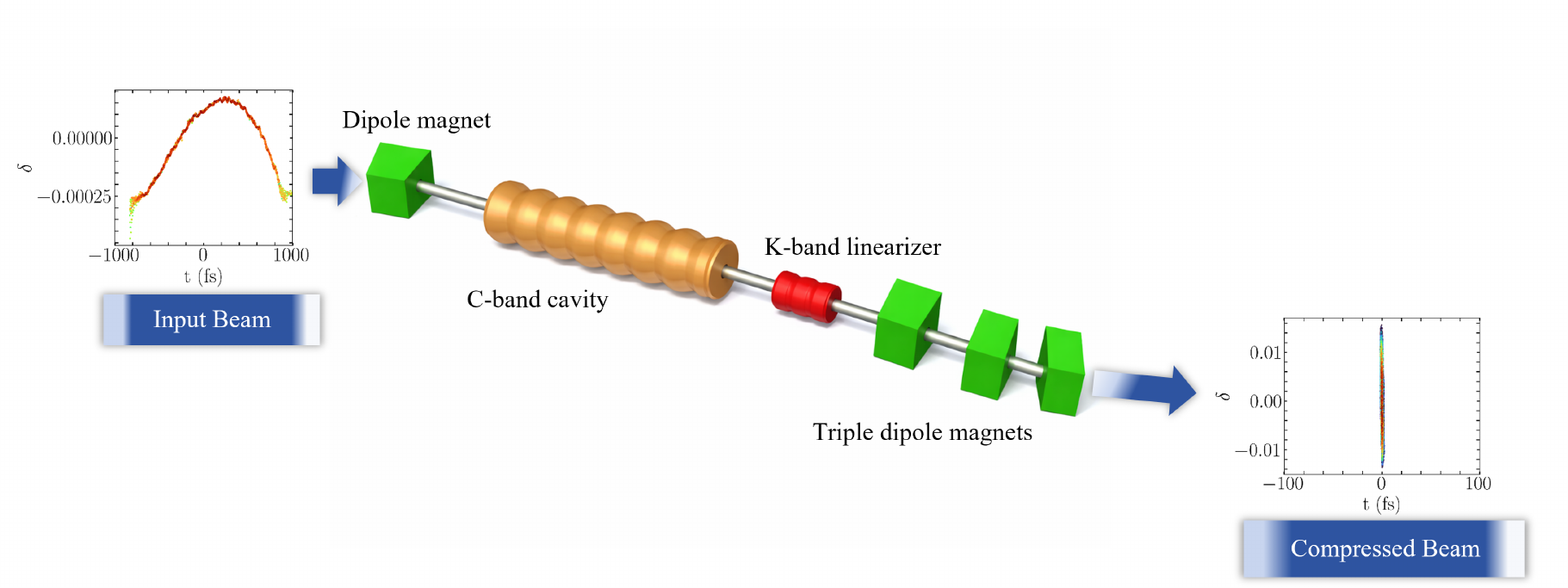}
 \caption{
Schematic layout of the proposed two-dimensional beam-compression scheme.
From left to right, a rectangular dipole introduces transverse--longitudinal
coupling, the C-band cavity provides the linear energy chirp, the K-band cavity
provides the second-order energy chirp needed to compensate the dominant
nonlinear contribution, and the three-dipole
dispersive section completes the compression. The insets show the
longitudinal phase spaces of the input and compressed beams, respectively.
}
\label{fig1}
\end{figure}

As schematically shown in Fig.~\ref{fig1}, the electron beam first passes through a rectangular dipole, where transverse-longitudinal correlations are introduced. The beam then enters a C-band RF cavity, which imposes a correlated energy chirp along the bunch. A K-band linearizer is subsequently used to compensate the second-order correlation between the beam energy and the longitudinal position. Finally, the beam is transported through a dispersive section consisting of three dipoles, where the bunch-length compression is achieved.

In the following, we analyze the proposed compression scheme from both linear and nonlinear perspectives. We first derive the basic compression mechanism using a linear transfer-map description, and then discuss the influence of higher-order terms on the final bunch length. A proof-of-principle compression demonstration based on an initially Gaussian electron beam is also presented.

\subsection{Linear case}
We first consider the idealized case in which the beam dynamics are described only by linear correlations among the phase-space coordinates. Since the compression mechanism is introduced in the bending plane, the electron motion is treated in two dimensions, corresponding to the horizontal and longitudinal degrees of freedom. The beam transport can therefore be described using a four-dimensional phase-space vector, $\mathbf{X} = \left( x, x', z, \delta \right)^{\mathrm{T}}$, where $x$ and $x^{\prime}$ are the transverse position and divergence in the bending plane, $z$ is the longitudinal position with respect to the reference particle, and $\delta$ is the relative energy deviation.

After the beam passes through the first rectangular dipole, the phase-space coordinates are transformed according to $\mathbf{X}_1 = R_b \mathbf{X}_0$, where $R_b$ is the transfer matrix of the dipole. When edge focusing is included and the incident beam is perpendicular to the entrance face of the dipole, the matrix $R_b$ 

\begin{equation}
R_b =
\left(
\begin{array}{cccc}
\sec\theta &
\displaystyle \frac{L\sin\theta}{\theta} &
0 &
\displaystyle \frac{L(1-\cos\theta)}{\theta}
\\[6pt]
0 &
\cos\theta &
0 &
\sin\theta
\\[6pt]
\tan\theta &
\displaystyle \frac{L(1-\cos\theta)}{\theta} &
1 &
\displaystyle \frac{L(\theta-\sin\theta)}{\theta}
\\[6pt]
0 & 0 & 0 & 1
\end{array}
\right) .
\label{eq:Rb}
\end{equation}
where \(L\) is the length of the dipole and \(\theta\) is the bending angle. The nonzero off-diagonal terms in \(R_b\) indicate that the rectangular dipole introduces transverse--longitudinal coupling. In particular, the elements \(R_{31}\) and \(R_{32}\) make the longitudinal coordinate depend on the incoming transverse coordinates, which is the key ingredient for the two-dimensional compression mechanism discussed below.

After the rectangular dipole, the electron beam receives an energy chirp from a C-band RF cavity. 
The RF wavelength is approximately \(53~\mathrm{mm}\), which is much longer than the bunch length under consideration. 
As a result, the sinusoidal RF waveform can be expanded around the reference phase, and only the first-order term is retained in the linear analysis. 
The energy change is then approximated as
\begin{equation}
\delta_2 = \delta_1 + h z_1 ,
\label{eq:rf_chirp}
\end{equation}
where \(h\) denotes the linear energy chirp introduced by the cavity. 
The corresponding transfer matrix is
\begin{equation}
\mathbf{X}_2 =
R_{\mathrm{RF}}\mathbf{X}_1,
\qquad
R_{\mathrm{RF}} =
\left(
\begin{array}{cccc}
1 & 0 & 0 & 0 \\
0 & 1 & 0 & 0 \\
0 & 0 & 1 & 0 \\
0 & 0 & h & 1
\end{array}
\right) .
\label{eq:rrf}
\end{equation}

The beam is then transported through a dispersive section composed of three dipoles, hereafter referred to as the TDS. This section couples the longitudinal position to the transverse motion, thereby linking the final bunch length to the transverse beam parameters. Its first-order transfer matrix is written as

\begin{equation}
R_{\mathrm{TDS}}
=
R_{b,3} R_{L,2} R_{b,2} R_{L,1} R_{b,1},
\label{eq:rtds}
\end{equation}
where \(R_{b,i}\) \((i=1,2,3)\) has the same form as the rectangular-dipole matrix in Eq.~\ref{eq:Rb}, using the length and bending angle of the corresponding dipole. The drift matrix \(R_L\) for a drift length \(L\) is given by

\begin{equation}
R_L =
\left(
\begin{array}{cccc}
1 & L & 0 & 0 \\
0 & 1 & 0 & 0 \\
0 & 0 & 1 & 0 \\
0 & 0 & 0 & 1
\end{array}
\right) .
\label{eq:RL}
\end{equation}

For convenience, the element in the \(i\)th row and \(j\)th column of the TDS transfer matrix \(R_{\mathrm{TDS}}\) is denoted by \(R_{\mathrm{TDS},ij}\).

The final longitudinal coordinate after the TDS is
\begin{equation}
\begin{aligned}
z_f
&=
\sum_{i=1}^{4} R_{\mathrm{TDS},3i}X_{2,i}  \\
&=
R_{\mathrm{TDS},31}x_{2}
+
R_{\mathrm{TDS},32}\cos\theta\, x_0'
+
\left(
R_{\mathrm{TDS},32}\sin\theta+R_{\mathrm{TDS},34}
\right)\delta_0
+
\left(
hR_{\mathrm{TDS},34}+1
\right)z_1 ,
\end{aligned}
\label{eq:zf_linear}
\end{equation}
where \(R_{\mathrm{TDS},3i}\) denotes the elements of the first-order transfer matrix of the TDS. 
To obtain the shortest compressed bunch length, the contributions from \(x_{2}\), \(\delta_0\), and \(z_1\) are cancelled by imposing
\begin{equation}
\begin{cases}
R_{\mathrm{TDS},31}=0, \\
hR_{\mathrm{TDS},34}+1=0, \\
R_{\mathrm{TDS},32}\sin\theta+R_{\mathrm{TDS},34}=0 .
\end{cases}
\label{eq:linear_cancellation_conditions}
\end{equation}
Under these conditions, the final longitudinal coordinate depends only on the initial transverse divergence \(x_0'\).
Substituting Eq.~\eqref{eq:linear_cancellation_conditions} into Eq.~\eqref{eq:zf_linear}, and assuming an initially Gaussian beam, the bunch length after compression becomes
\begin{equation}
\sigma_{z,f}
=
\left|
\frac{1}{h\tan\theta}
\right|
\sigma_{x_0'}
=
\left|
\frac{1}{h\tan\theta}
\right|
\sqrt{
\frac{\gamma_{x,0}\varepsilon_n}{\gamma}
},
\label{eq:sigma_zf_linear}
\end{equation}
where \(\gamma_{x,0}\) is the initial horizontal Twiss parameter, \(\gamma\) is the relativistic factor, and \(\varepsilon_n\) is the normalized horizontal emittance.
The prefactor of \(\sigma_{x_0'}\) in Eq.~\eqref{eq:sigma_zf_linear} maps the initial transverse divergence into the final longitudinal coordinate and therefore sets the achievable compressed bunch length. We define this prefactor as the two-dimensional compression parameter,
\begin{equation}
C_{\mathrm{2D}}
=
\left|
\frac{1}{h\tan\theta}
\right| .
\label{eq:c2d}
\end{equation}
Equation~\ref{eq:sigma_zf_linear} shows a key feature of the proposed two-dimensional compression scheme: after the linear cancellation conditions are satisfied, the final bunch length is governed by the transverse geometric emittance rather than by the initial energy spread. This makes it possible to exploit the ultralow transverse emittance of modern electron beams for sub-femtosecond bunch generation.

It is worth noting that, in the limit of a small bending angle \(\theta\) for the first dipole, and if the RF cavity is replaced by a laser modulator, the present configuration reduces to the ADM scheme~\cite{Feng2017}. ADM has been extensively studied in storage-ring light sources as a method for generating microbunching and coherent radiation~\cite{Jiang2022,PhysRevAccelBeams.23.110701,rlt4-8bn9}. In ADM, the transverse--longitudinal coupling introduced by a small dipole or angular-dispersion section is combined with laser-induced energy modulation and a downstream dispersive section, leading to density modulation at higher harmonics.

Although the underlying coupling mechanism is closely related, the objective of the present work is different. Instead of producing periodic microbunching for coherent harmonic radiation, the proposed scheme uses the same transverse--longitudinal coupling principle to compress the entire electron bunch into the sub-femtosecond regime.

The downstream dispersive section also differs from that used in ADM. 
Here, the dispersive section is composed of three dipoles. 
Although its geometry is similar to that of a chicane, it is not constrained to be achromatic or coaxial. 
This chicane-like configuration provides a significant advantage over the two-dipole dogleg commonly used in ADM: it weakens the direct constraint between the RF chirp \(h\) and the first-dipole bending angle \(\theta\). 
Consequently, for a given value of \(h\), the bending angle can be optimized over a wider range, providing additional freedom to minimize \(C_{\mathrm{2D}}\).

To quantify this advantage, we perform a comparative optimization for dogleg and TDS. 
For different values of \(h\) and \(\theta\), we solve for all beamline parameters that satisfy the linear cancellation conditions in Eq.~\eqref{eq:linear_cancellation_conditions}. 
{For each solution, we calculate the compression parameter
\(C_{\mathrm{2D}}\) and the corresponding linear compression ratio using the beam parameters listed in Table~\ref{tab:beam_parameters}.}

\begin{figure}
 \centering
        \includegraphics[width=1\textwidth]{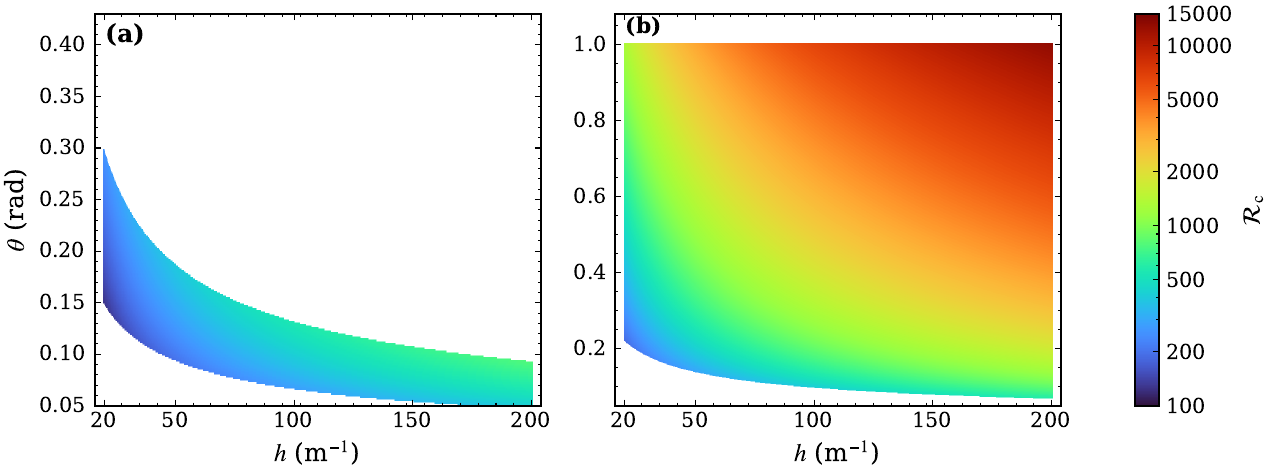}
 \caption{
{Comparison of the linear compression ratio
\(\mathcal{R}_{\mathrm{c}}=\sigma_{z,0}/\sigma_{z,f}\) for the dogleg (a)
and TDS (b) at different RF chirps \(h\) and first-dipole bending angles
\(\theta\).  The ratio is evaluated using the beam parameters in
Table~\ref{tab:beam_parameters}, including \(\gamma_{x,0}=0.01~\mathrm{m}^{-1}\).}
The length of each dipole is fixed at \(0.2~\mathrm{m}\). 
For the dogleg, the dipole separation and bending angle are treated as free variables. For the TDS, the dipole separation is fixed at \(0.3~\mathrm{m}\), and the bending angles are varied. For a fair comparison, the minimum dipole separation in the dogleg is also set to \(0.3~\mathrm{m}\).
The white region corresponds to parameter ranges where the linear compression conditions cannot be satisfied. 
}
\label{fig:dogleg_tds_comparison}
\label{fig2}
\end{figure}

{Figure~\ref{fig:dogleg_tds_comparison} compares the linear
compression ratios achievable with the dogleg and TDS configurations.} For the
dogleg, solutions cease to exist above an \(h\)-dependent limit on \(\theta\),
corresponding to the white region in which the linear cancellation conditions
in Eq.~\eqref{eq:linear_cancellation_conditions} cannot be satisfied. In
contrast, the TDS admits valid solutions over a substantially wider range of
bending angles.

{The compression advantage of the TDS is also evident in
Fig.~\ref{fig:dogleg_tds_comparison}, where it achieves higher linear
compression ratios within the feasible region and provides greater flexibility
for optimizing two-dimensional beam compression.}

\subsection{Nonlinear compensation}

The linear analysis in the previous subsection establishes the basic conditions for two-dimensional beam compression and shows that, once the linear cancellation conditions are satisfied, the final bunch length is determined by the transverse geometric emittance. 
However, this first-order description becomes insufficient as the bending angle \(\theta\) increases. 
A larger bending angle enhances the transverse--longitudinal coupling and can reduce the compression parameter, but it also strengthens nonlinear transport effects. 
These higher-order contributions can broaden the final longitudinal distribution and ultimately limit the minimum achievable bunch length.

Therefore, for sub-femtosecond compression, nonlinear compensation is required. 
In deriving Eq.~\ref{eq:sigma_zf_linear}, all higher-order contributions were neglected. When the dominant nonlinear contribution is retained, the final longitudinal coordinate can be approximately written as
\begin{equation}
z_f =
\sum_{i=1}^{4} R_{\mathrm{TDS},3i} X_{2,i}
+
\sum_{i\leq j} T_{3ij} X_{2,i} X_{2,j}
\simeq
\sum_{i=1}^{4} R_{\mathrm{TDS},3i} X_{2,i}
+
T_{344} h^2 z_1^2 ,
\label{eq:zf_nonlinear}
\end{equation}
Here, \(T_{3ij}\) are the second-order path-length transport coefficients of the TDS. The term \(T_{344}h^2z_1^2\) represents the dominant nonlinear contribution after the RF cavity, arising from the quadratic dependence of the final longitudinal position on the energy deviation induced by the chirp.

To compensate the dominant second-order contribution, we introduce a K-band linearizer downstream of the C-band cavity. 
The K-band linearizer provides a controllable second-order energy chirp \(h_2\), which can be used to cancel the quadratic dependence of the final longitudinal coordinate on the longitudinal position.

After the C-band cavity and the K-band linearizer, the relative energy deviation of an electron can be written as
\begin{equation}
\delta_2
=
\delta_1
+
V_1 \sin \left( k_1 z_1+\phi_1 \right)
+
V_2 \cos \left(3 k_1 z_1 \right),
\label{eq:delta_rf_kband}
\end{equation}
where \(V_1\) and \(V_2\) denote the normalized voltage amplitudes of the C-band cavity and the K-band linearizer, respectively, \(k_1\) is the wave number of the C-band RF field, and \(\phi_1\) is the phase of the C-band cavity. 
The voltage amplitudes are chosen to satisfy
\begin{equation}
V_1 \sin\phi_1 + V_2 = 0 ,
\label{eq:rf_energy_offset}
\end{equation}
so that the reference particle has no net energy change after the two RF cavities.

Since the RF wavelength is much longer than the bunch length, namely \(3 k_1 z_1 \ll 1\), Eq.~\ref{eq:delta_rf_kband} can be expanded around the reference particle. 
Keeping terms up to second order in \(z_1\), we obtain
\begin{equation}
\delta_2
=
\delta_1
+
h z_1
+
h_2 z_1^2 ,
\label{eq:delta_h_h2}
\end{equation}
where \(h\) and \(h_2\) are the linear and second-order energy chirps, respectively.

Substituting Eq.~\ref{eq:delta_h_h2} into Eq.~\ref{eq:zf_nonlinear}, and retaining only the dominant second-order contribution, gives
\begin{equation}
z_f
\simeq
\sum_{i=1}^{4} R_{\mathrm{TDS},3i} X_{2,i}
+
\left(
T_{344} h^2
+
R_{\mathrm{TDS},34} h_2
\right) z_1^2 .
\label{eq:zf_second_order_compensation}
\end{equation}

Therefore, the dominant quadratic term can be cancelled by choosing
\begin{equation}
h_2
=
-
\frac{T_{344}}{R_{\mathrm{TDS},34}}
h^2 .
\label{eq:h2_compensation}
\end{equation}

In practice, for a given linear chirp \(h\), the compensation condition in Eq.~\ref{eq:h2_compensation} does not need to be implemented by explicitly calculating the matrix elements. 
Instead, the RF parameters can be optimized more directly. Specifically, the C-band voltage is expressed as
\begin{equation}
V_1 =
\frac{h}{k_1 \cos\phi_1},
\label{eq:V1_from_h}
\end{equation}
and the K-band voltage is then determined from Eq.~\ref{eq:rf_energy_offset}. 
Finally, by adjusting the phase \(\phi_1\), the resulting second-order chirp can be made to satisfy the compensation condition in Eq.~\ref{eq:h2_compensation}. With this approach, one only needs to scan the voltage and phase of the C-band cavity, while the K-band voltage is determined accordingly. Further details are provided in Section~4.

To demonstrate the basic feasibility of the proposed compression scheme, we carry out a proof-of-principle simulation using the electron-beam parameters listed in Table~\ref{tab:beam_parameters}. 
The purpose of this example is not to represent a fully optimized accelerator design, but to verify that the linear compression conditions and nonlinear compensation strategy discussed above can lead to sub-femtosecond bunch generation with realistic beam parameters. 

\begin{table}
\caption{
Main parameters used for the proof-of-principle demonstration.
}
\centering
\begin{tabular}{l c}
\hline
Parameter & Value \\
\hline
Beam energy & \(200~\mathrm{MeV}\) \\
Normalized emittance & \(0.08~\mu\mathrm{m}\) \\
Rms energy spread & \(2\times 10^{-4}\) \\
Bunch length & {\(60~\mu\mathrm{m}\)} \\
Bunch charge & \(4~\mathrm{pC}\) \\
\hline
\end{tabular}
\label{tab:beam_parameters}
\end{table}

The beamline parameters are first optimized by solving the linear cancellation conditions in Eq.~\eqref{eq:linear_cancellation_conditions}. 
The main optimized parameters are listed in Table~\ref{tab:optimized_parameters}. 

\begin{table}[t]
\caption{
Main parameters of the compression beamline.
}
\centering
\begin{tabular}{l c}
\hline
Parameter & Value \\
\hline
\multicolumn{2}{l}{\textit{First dipole}} \\
Length & \(0.3~\mathrm{m}\) \\
Bending angle & \(0.382~\mathrm{rad}\) \\
\multicolumn{2}{l}{\textit{C-band RF}} \\
\(V_1\) & \(56.15~\mathrm{MV}\) \\
\(\phi_1\) & \(0.095~\mathrm{rad}\) \\
\multicolumn{2}{l}{\textit{K-band RF}} \\
\(V_2\) & \(-5.33~\mathrm{MV}\) \\
\(\phi_2\) & \(\pi/2\) \\
\multicolumn{2}{l}{\textit{TDS}} \\
Dipole 1 length & \(0.2~\mathrm{m}\) \\
Dipole 1 bending angle & \(0.112~\mathrm{rad}\) \\
Dipole 2 length & \(0.2~\mathrm{m}\) \\
Dipole 2 bending angle & \(-0.352~\mathrm{rad}\) \\
Dipole 3 length & \(0.2~\mathrm{m}\) \\
Dipole 3 bending angle & \(0.250~\mathrm{rad}\) \\
All drift lengths & \(0.3~\mathrm{m}\) \\
\hline
\end{tabular}
\label{tab:optimized_parameters}
\end{table}

\begin{figure}[t]
 \centering
        \includegraphics[width=0.32\textwidth]{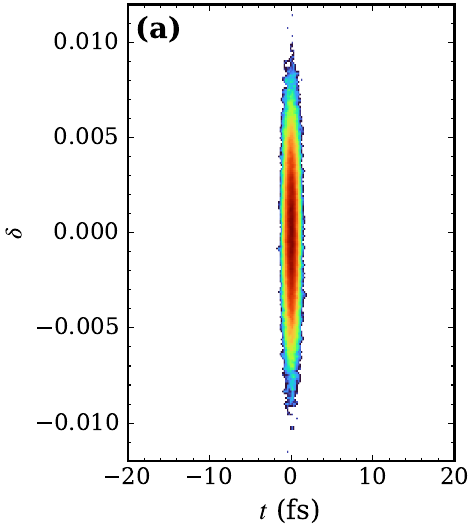}\hfill
        \includegraphics[width=0.32\textwidth]{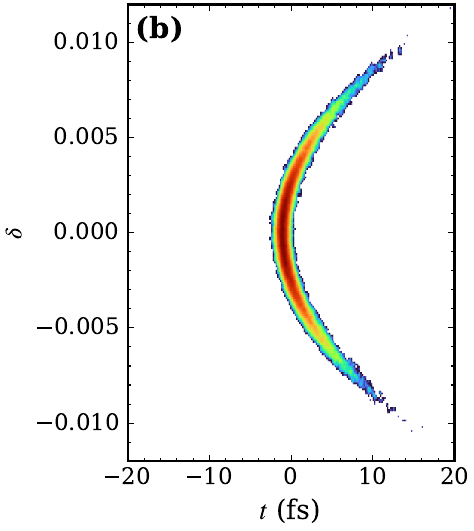}\hfill
        \includegraphics[width=0.32\textwidth]{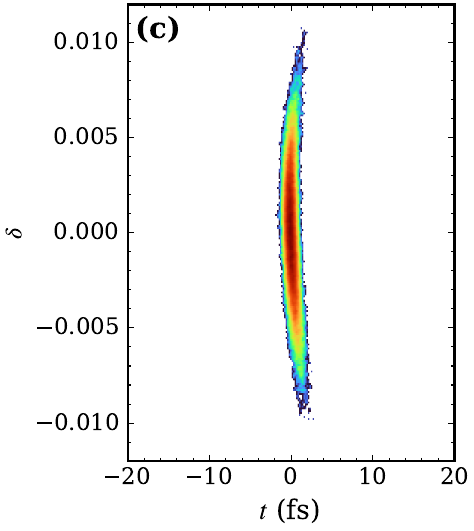}

 \caption{
Verification of the two-dimensional compression scheme in the zero-current
limit using ELEGANT simulations. The panels show the final longitudinal phase
space \((t,\delta)\), with color indicating particle density. (a) Ideal linear
compression; (b) compression including higher-order transport effects without
second-order compensation; and (c) compression after second-order compensation
with the K-band linearizer.
}
\label{fig:linear_nonlinear_compensation}
\label{fig3}
\end{figure}

Using the ELEGANT code~\cite{osti_761286}, we further verify the compression performance in the zero-current limit. 
Three cases are considered: the ideal linear compression, the compression including higher-order transport effects without second-order compensation, and the compression with second-order compensation. 
The results are shown in Fig.~\ref{fig:linear_nonlinear_compensation}.

In the ideal linear case, the compressed bunch length reaches \(388~\mathrm{as}\). 
When higher-order effects are included without compensation, the bunch length increases to \(1.63~\mathrm{fs}\), indicating that nonlinear transport terms can significantly degrade the compression performance. 
After compensating the dominant second-order term with the K-band linearizer, the compressed bunch length is reduced to \(445~\mathrm{as}\), which is very close to the result obtained under the linear approximation. 
This agreement confirms that the dominant higher-order effect is effectively controlled by the proposed nonlinear compensation scheme.

\section{Dependence of the compressed bunch length on charge and beam energy}
\label{sec:length_charge_energy}

{When a finite bunch charge is considered, collective effects may limit the
minimum achievable bunch length. We first consider the ranges
\(0\leq Q\leq 50~\mathrm{pC}\) and
\(100\leq E\leq 600~\mathrm{MeV}\), for which ISR is expected to be weak. Simulations including only ISR confirm
that its effect on the final bunch length is negligible in this range.
We therefore first focus on CSR and SC, while retaining ISR in the simulations for completeness. }

We next consider CSR and examine whether multidimensional effects must be included.
According to the Derbenev criterion, the one-dimensional CSR approximation is expected to be valid when
\begin{equation}
\frac{\sigma_x}{\left(\rho \sigma_z^2\right)^{1/3}} \ll 1 ,
\label{eq:derbenev_criterion}
\end{equation}
where \(\sigma_x\) is the transverse beam size in the bending plane, \(\rho\) is the bending radius, and \(\sigma_z\) is the bunch length. 
For most of the compression beamline considered here, this condition is well satisfied. 
The only exception occurs in the final dipole, where the bunch length becomes very short and the Derbenev parameter can reach values on the order of \(10\). 
This suggests that transverse CSR effects may become non-negligible in the last dipole.

\begin{figure}
 \centering
        \includegraphics[width=0.5\textwidth]{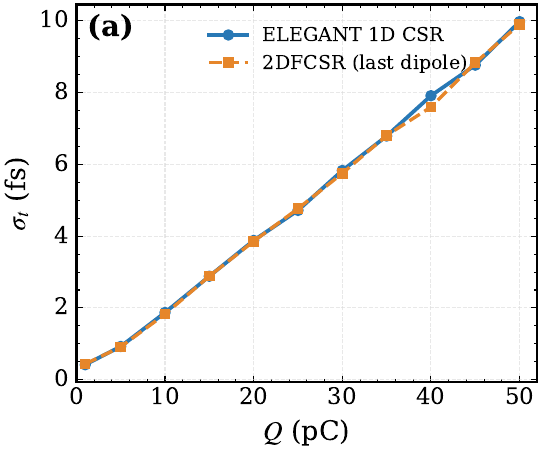}\hfill
        \includegraphics[width=0.5\textwidth]{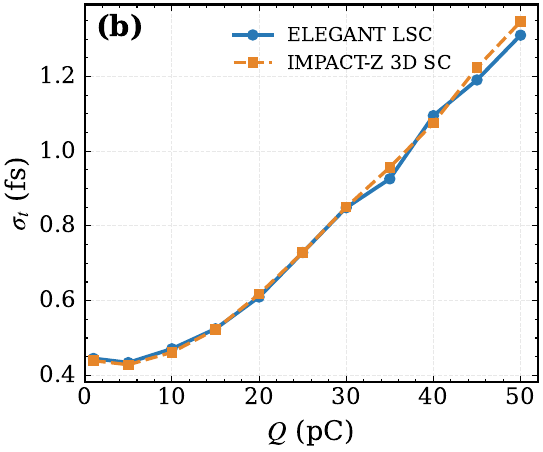}
 \caption{
Comparison of collective-effect models as a function of bunch charge at
\(200~\mathrm{MeV}\).
(a) The blue solid curve with circles is the ELEGANT one-dimensional CSR
result, and the orange dashed curve with squares is the 2D-CSR calculation for
the last dipole. (b) The blue solid curve with circles is the ELEGANT
one-dimensional longitudinal-space-charge result, and the orange dashed curve
with squares is the IMPACT-Z three-dimensional space-charge result.
}
\label{fig:collective_effects}
\label{fig4}
\end{figure}

To evaluate the possible contribution of multidimensional CSR in this region, we compare one-dimensional CSR simulations with calculations performed using the 2DFCSR code~\cite{tang:fls2023-th3d4}. 
The results are shown in Fig.~\ref{fig:collective_effects}. 
For bunch charges in the range from \(0\) to \(50~\mathrm{pC}\), the final bunch lengths obtained with the one-dimensional and two-dimensional CSR models are found to be nearly identical. 
This indicates that, for the present beamline and parameter range, the influence of two-dimensional CSR on the compressed bunch length is negligible. 
The CSR-induced bunch-length increase is therefore mainly determined by the one-dimensional CSR effect.

Finally, we examine the influence of space charge. 
To determine whether a simplified longitudinal model is sufficient, we compare simulations including only one-dimensional longitudinal space charge with those including full three-dimensional space charge~\cite{10.1145/331532.331587}. 
As shown in Fig.~\ref{fig:collective_effects}, the two models give very similar final bunch lengths over the charge range from \(0\) to \(50~\mathrm{pC}\). 
This comparison shows that the one-dimensional longitudinal space-charge model captures the dominant contribution of SC to the bunch-length degradation in the present case.

It is also found that, within the charge range below \(50~\mathrm{pC}\), the bunch-length degradation is dominated by CSR rather than SC. 
When the bunch charge exceeds approximately \(5~\mathrm{pC}\), CSR already lengthens the compressed bunch to nearly \(1~\mathrm{fs}\). 
In contrast, the SC effect produces only a negligible change in the final bunch length at the same charge level. 
This behavior indicates that CSR is the primary collective effect limiting the compression performance in the present beamline.

\begin{figure}
 \centering
        \includegraphics[height=0.23\textheight]{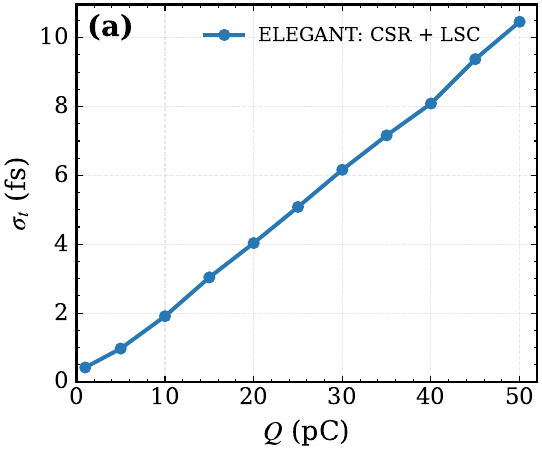}\hfill
        \includegraphics[height=0.23\textheight]{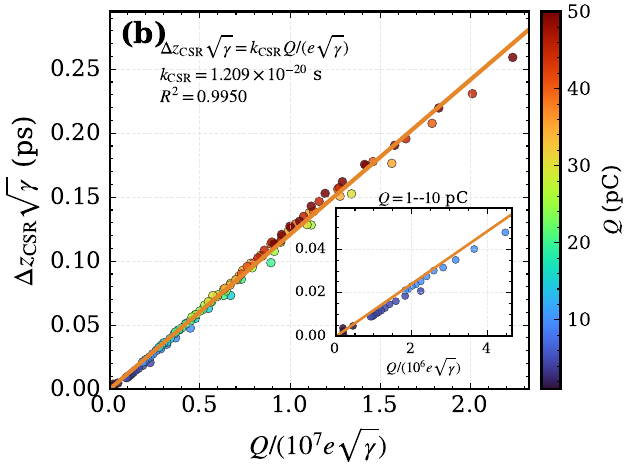}
 \caption{{
(a) Final bunch length as a function of bunch charge at
\(200~\mathrm{MeV}\), calculated with ELEGANT including one-dimensional CSR and
longitudinal space charge. The blue circles are the simulation results, and
the connecting line is a guide to the eye. (b) Normalized effective CSR
contribution, \(\Delta z_{\mathrm{CSR}}\sqrt{\gamma}\), plotted against
\(Q/(e\sqrt{\gamma})\). The symbols are Gaussian-beam simulations at different
bunch charges and beam energies, their colors encode bunch charge, and the
orange solid line is the fit based on Eq.~\ref{eq:sigma_csr_scaling}. The inset
enlarges the low-charge range \(1\leq Q\leq10~\mathrm{pC}\).
}}
\label{fig:bunch_length_charge}
\label{fig:scaling_fit}
\label{fig5}
\end{figure}

To evaluate the combined influence of CSR and SC, we perform ELEGANT simulations including both one-dimensional CSR and longitudinal space charge. 
The bunch charge is varied from \(0\) to \(50~\mathrm{pC}\), while the beamline parameters are kept fixed. 
As shown in Fig.~\ref{fig:bunch_length_charge}, the final bunch length increases approximately linearly with bunch charge. 
This nearly linear dependence provides a useful empirical scaling for estimating the charge limit of the proposed two-dimensional compression scheme.

{
According to Eq.~\ref{eq:sigma_zf_linear}, the ideal bunch length in the
absence of collective effects scales as \(1/\sqrt{\gamma}\). We therefore
define the normalized ideal bunch length as
\begin{equation}
\sigma_{n,0}
=
\sqrt{\gamma}\,\sigma_{t,\mathrm{ideal}}
=
\frac{1}{c|h\tan\theta|}
\sqrt{\gamma_{x,0}\varepsilon_n},
\label{eq:normalized_bunch_length}
\end{equation}
where \(c\) is the speed of light in vacuum. For fixed compression optics,
\(\sigma_{n,0}\) is independent of the beam energy.
\par
}

{
To quantify the CSR-induced degradation of the bunch length, we define an effective CSR contribution by
\begin{equation}
\Delta z_{\mathrm{CSR}}
\equiv
\sqrt{
\sigma_{t,\mathrm{noISR}}^{\,2}
-
\sigma_{t,\mathrm{ideal}}^{\,2}
},
\label{eq:effective_csr_contribution}
\end{equation}
where \(\sigma_{t,\mathrm{noISR}}\) is obtained with CSR and SC included but
ISR disabled. Here, \(\Delta z_{\mathrm{CSR}}\) denotes an effective rms
longitudinal contribution expressed in units of time.
\par
}

{
As the beam passes through the TDS, the CSR interaction is strongest in the
final dipole, where the bunch has already undergone substantial compression.
The following analysis therefore focuses on the CSR effects in this dipole.
According to the CSR
kick model~\cite{PhysRevSTAB.17.060701,PhysRevAccelBeams.28.024402}, the
induced coordinate perturbations scale as
\[
\Delta x'
\propto
\frac{Q}{\gamma\sigma_z^{4/3}}g_{x'},
\qquad
\Delta\delta
\propto
\frac{Q}{\gamma\sigma_z^{4/3}}g_{\delta}.
\]
Here, \(Q\) is the bunch charge, and \(\sigma_z\) is the bunch length at the
entrance to the final dipole. In the present scaling estimate, \(\sigma_z\) is
approximated by the initial bunch length. The functions \(g_{x'}\) and \(g_{\delta}\) are correction
factors that describe the bunch-length dependence of the CSR-induced angular
and energy kicks, respectively. When the final bunch length is much shorter
than the initial bunch length, these correction factors are determined mainly
by the beamline geometry and depend only weakly on the initial beam parameters.

For the fixed initial Gaussian bunch length and the beamline considered here, the
dependence on the initial bunch length and beamline geometry can be absorbed
into a coefficient \(k_{\mathrm{CSR}}\). The effective CSR contribution can
then be written as
\begin{equation}
\Delta z_{\mathrm{CSR}}
=
k_{\mathrm{CSR}}\frac{Q}{e\gamma}.
\label{eq:sigma_csr_scaling}
\end{equation}
Thus, within the parameter range considered in the present study,
\(k_{\mathrm{CSR}}\) is independent of \(Q\) and \(\gamma\) (or,
equivalently, the beam energy \(E\)), although it is not a universal constant.
\par
}

{
The final bunch length when ISR is negligible is consequently
\begin{equation}
\sigma_{t,\mathrm{nonISR}}
=
\sqrt{
\sigma_{t,\mathrm{ideal}}^{\,2}
+
\Delta z_{\mathrm{CSR}}^{\,2}
}.
\label{eq:sigma_total_csr}
\end{equation}
Equivalently, the normalized final bunch length is
\begin{equation}
\sigma_{n,f}
=
\sqrt{
\sigma_{n,0}^{\,2}
+
\left(
k_{\mathrm{CSR}}\frac{Q}{e\sqrt{\gamma}}
\right)^2
}.
\label{eq:normalized_final_length}
\end{equation}
A fit to the Gaussian-beam simulations gives
\begin{equation}
k_{\mathrm{CSR}}
=
1.209\times10^{-20}~\mathrm{s},
\label{eq:k_value}
\end{equation}
with a coefficient of determination of \(R^2=0.995\). As shown in Fig.~\ref{fig:scaling_fit}, the simulated
effective CSR contributions are well reproduced by
Eq.~\ref{eq:sigma_csr_scaling}.
\par
}

{
Equation~\ref{eq:normalized_final_length} is valid only when ISR produces a
negligible correction. Increasing the beam energy suppresses the CSR term,
but the ISR-induced energy spread increases rapidly with energy and eventually
limits the compression.
\par
}

{
The effective ISR contribution is described by~\cite{Chao2006}
\begin{equation}
\Delta z_{\mathrm{ISR}}
=
k_{\mathrm{ISR}}\gamma^{5/2},
\label{eq:isr_contribution}
\end{equation}

\par
}

{

The complete bunch-length model is therefore
\begin{equation}
\sigma_{t,\mathrm{full}}
=
\sqrt{
\sigma_{t,\mathrm{ideal}}^{\,2}
+
\Delta z_{\mathrm{CSR}}^{\,2}
+
\Delta z_{\mathrm{ISR}}^{\,2}
}.
\label{eq:complete_bunch_length}
\end{equation}
\par

To examine this transition, we extend the Gaussian-beam scan to
\(1\leq Q\leq60~\mathrm{pC}\) and
\(100\leq E\leq2500~\mathrm{MeV}\), while retaining the same initial beam
parameters and compressor geometry. The RF settings are adjusted at each
energy to preserve the required compression conditions. 

Fitting Eq.~\ref{eq:isr_contribution} to the scan data yields
\begin{equation}
k_{\mathrm{ISR}}
=
2.628\times10^{-24}~\mathrm{s}.
\label{eq:k_isr_value}
\end{equation}

}

\begin{figure}[ht]
\centering
\includegraphics[width=\textwidth]{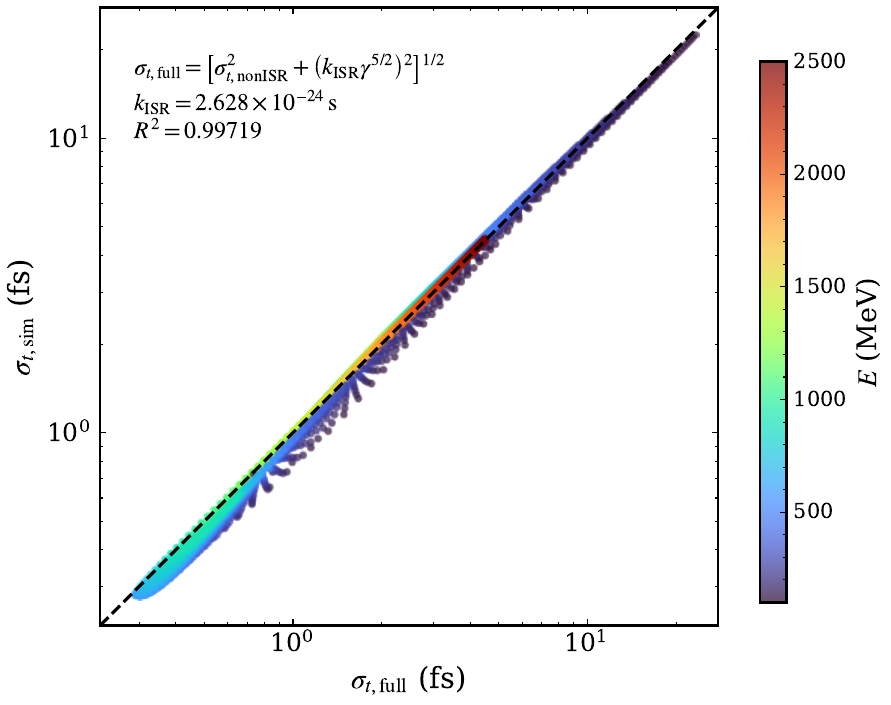}
\caption{{
Validation of the complete bunch-length model in
Eq.~\ref{eq:complete_bunch_length} against the all-effects Gaussian-beam
simulations. The horizontal axis shows the theoretical prediction
\(\sigma_{t,\mathrm{full}}\), and the vertical axis shows the simulated bunch
length \(\sigma_{t,\mathrm{sim}}\). Each point represents a \(Q\)--\(E\)
combination in the extended parameter scan and is colored according to the
beam energy \(E\). The black dashed line denotes exact agreement between the
theory and simulation. The annotation gives the fitted value of
\(k_{\mathrm{ISR}}\) and \(R^2=0.99719\).
}}
\label{fig:two_region_validation}
\end{figure}

{
As shown in Fig.~\ref{fig:two_region_validation}, the predictions obtained
from the fitted model agree closely with the simulation data, with
\(R^2=0.997\).
\par
}

\begin{figure}[ht]
\centering
\includegraphics[width=0.94\textwidth]{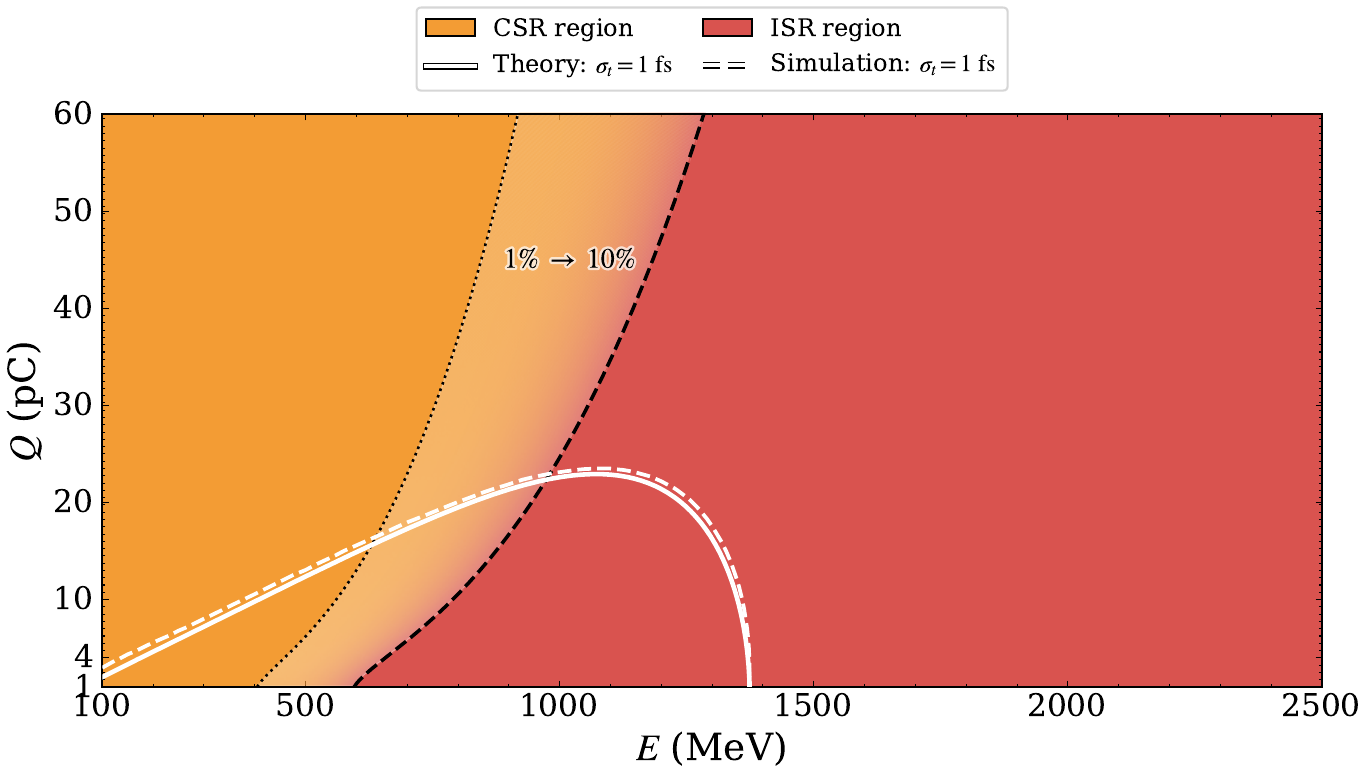}
\caption{{
Operating map in the bunch-charge--beam-energy (\(Q\)--\(E\)) plane for the
Gaussian input beam. The background color indicates the transition from the
CSR-dominated regime (orange) to the ISR-dominated regime (red). The black
dotted and dashed curves mark the regime boundaries obtained with ISR
tolerances of \(1\%\) and \(10\%\), respectively. As the tolerance increases,
the boundary shifts toward the ISR-dominated regime, enlarging the parameter
range described by the CSR-only model. The solid white curve is the
\(\sigma_t=1~\mathrm{fs}\) contour predicted by
Eq.~\ref{eq:complete_bunch_length}, whereas the dashed white curve is obtained
from simulations. The region below each white contour corresponds to
sub-femtosecond compression.
}}
\label{fig:qe_operating_map}
\end{figure}

{
Using Eqs.~\ref{eq:sigma_total_csr} and~\ref{eq:complete_bunch_length}, we
divide the scanned \(Q\)--\(E\) parameter space into two regimes: a
CSR-dominated regime and an ISR-dominated regime. The CSR-dominated regime is
defined by the following criterion:
\[
\frac{\sigma_{t,\mathrm{full}}}
{\sigma_{t,\mathrm{nonISR}}}
-1
\leq 10\%.
\]
Equation~\ref{eq:sigma_total_csr} is used in this region; above the boundary,
the ISR correction in Eq.~\ref{eq:complete_bunch_length} is required.
Choosing a different tolerance would shift the boundary quantitatively, but
would not change the existence of the two operating regions or their
characteristic dependence on beam energy, as shown in
Fig.~\ref{fig:qe_operating_map}. The figure also shows the boundaries obtained
for tolerances ranging from \(1\%\) to \(10\%\). As the tolerance increases,
the boundary shifts toward the ISR-dominated regime, thereby enlarging the
CSR-dominated regime.
}

{
Figure~\ref{fig:qe_operating_map} also shows the contours corresponding to a
compressed bunch length of \(1~\mathrm{fs}\). The solid white contour is
calculated using Eq.~\ref{eq:complete_bunch_length}, whereas the dashed white
contour is obtained from simulations. Their close agreement further validates
the theoretical model. These contours provide a direct estimate of the
operating capability of the present compressor. For the specified initial Gaussian beam,
bending angles, and beamline geometry, sub-femtosecond operation is expected
below the solid white curve in Fig.~\ref{fig:qe_operating_map}. The theoretical
boundary reaches a maximum bunch charge of approximately \(22.9~\mathrm{pC}\)
near \(1.07~\mathrm{GeV}\). In the limit of vanishing charge, ISR sets an
upper beam-energy boundary of approximately \(1.37~\mathrm{GeV}\). These
values characterize the present compressor configuration rather than a
universal limit of two-dimensional beam compression.
\par
}

\clearpage

\section{Sub-femtosecond beam generation}
\label{sec:subfs_generation}

{
The operating map obtained in Sec.~\ref{sec:length_charge_energy} predicts
that the \(200~\mathrm{MeV}\), \(4~\mathrm{pC}\) working point lies below the
\(1~\mathrm{fs}\) boundary. It is therefore expected to support
sub-femtosecond compression while remaining compatible with a compact
accelerator. We select this point for a detailed start-to-end validation of
the proposed scheme.
\par
}

The pre-injector configuration is similar to those adopted in compact C-band photocathode-linac designs~\cite{Jiang2026}. A \(4~\mathrm{pC}\) electron bunch is generated from a photocathode driven by a \(266~\mathrm{nm}\) laser. The initial kinetic energy of the emitted electrons is sampled from a Fermi--Dirac distribution, with an average energy of \(0.287~\mathrm{eV}\). After acceleration in a 3.6 cell photocathode RF gun, the beam reaches an energy of \(7.3~\mathrm{MeV}\) at the gun exit.

The beam is then transported through a solenoid section for emittance compensation, with a main solenoid field of \(0.3~\mathrm{T}\). After the gun section, the beam enters a \(2~\mathrm{m}\)-long travelling-wave accelerating structure consisting of 114 cells. The beam is subsequently accelerated by three travelling-wave structures with peak gradients of \(26\), \(45\), and \(45~\mathrm{MV/m}\), respectively. At the exit of the accelerating section, the beam energy reaches \(200~\mathrm{MeV}\), while the normalized emittance is maintained at approximately \(0.0866~\mathrm{\mu m}\). The injector and accelerating sections described above are simulated using the ASTRA code~\cite{Floettmann1997}. The resulting ASTRA-generated distribution is used as the input to the downstream compressor tracking. Hereafter, beam distributions obtained from this injector-to-compressor workflow are referred to as start-to-end beams. The main parameters are listed in Table~\ref{tab:start_to_end_parameters}.

\setcounter{table}{2}
\begin{table}[t]
\caption{
Main beam parameters obtained from ASTRA simulations.
}
\centering
\begin{tabular}{l c}
\hline
Parameter & Value \\
\hline
Beam energy & \(200~\mathrm{MeV}\) \\
Normalized emittance & \(0.0866~\mu\mathrm{m}\) \\
Rms energy spread & \(1.4\times 10^{-4}\) \\
Bunch length & \(138~\mu\mathrm{m}\) \\
Bunch charge & \(4~\mathrm{pC}\) \\
\hline
\end{tabular}
\label{tab:start_to_end_parameters}
\end{table}

After acceleration, the beam is focused by four quadrupole magnets and transported into the transverse--longitudinal coupled compression section. 
The dipole configuration of the compression section is kept the same as that used in the Gaussian-beam study in the previous section. 
However, because the start-to-end beam distribution contains realistic correlations and nonlinearities from the injector and accelerating sections, the voltages and phases of the C-band cavity and the K-band linearizer must be re-optimized.

In the optimization procedure, the K-band linearizer is first turned off. 
The phase of the C-band cavity is set to the zero-crossing phase, so that it provides the required linear energy chirp without changing the reference beam energy. 
The corresponding optimized voltage of the C-band cavity is denoted by \(V_{1,0}\). 
This voltage is then used as the initial reference for the subsequent joint optimization of the C-band cavity and the K-band linearizer.

\begin{figure}[t]
 \centering
        \includegraphics[width=0.88\textwidth]{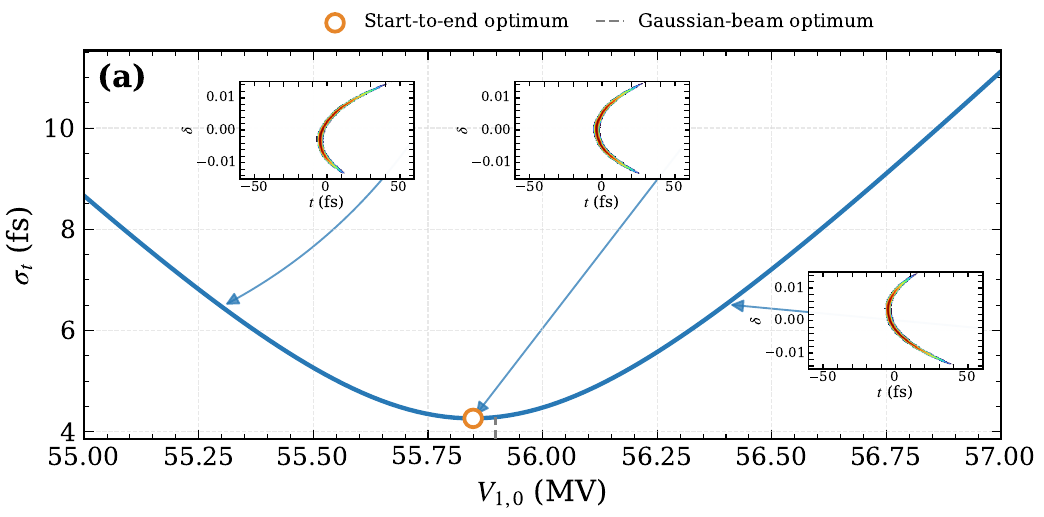}\\[1mm]
        \includegraphics[width=0.88\textwidth]{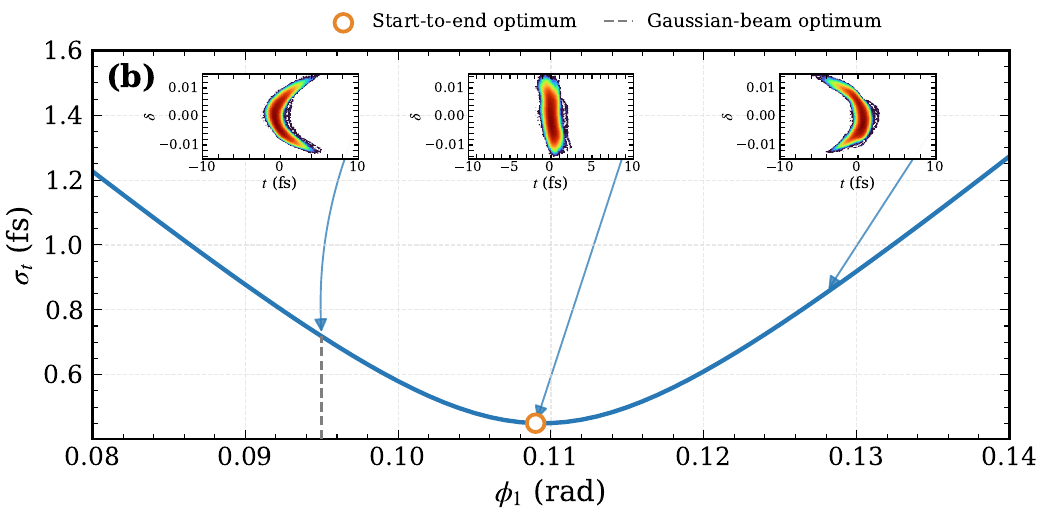}
 \caption{
RF optimization for the start-to-end beam. (a) Final bunch length versus
the C-band zero-crossing voltage \(V_{1,0}\), with the K-band linearizer turned
off. (b) Final bunch length versus the C-band phase \(\phi_1\), with the
K-band linearizer turned on and the two cavity voltages constrained by
Eqs.~\ref{eq:V1_phase_scan} and \ref{eq:V2_phase_scan}. In both panels, the
blue curve is the start-to-end scan, the orange open circle marks its optimum,
and the grey dashed marker gives the Gaussian-beam optimum. The insets show the
final longitudinal phase spaces at the settings indicated by the blue arrows.
}
\label{fig6}
\end{figure}

Figure~\ref{fig6} shows the optimization of the C-band cavity voltage \(V_{1,0}\) for the start-to-end beam distribution. 
Compared with the ideal Gaussian beam, the optimal value of \(V_{1,0}\) for the start-to-end beam is slightly smaller. 
This indicates that the linear compression condition obtained from the Gaussian-beam model remains a good initial estimate, but a small re-optimization is required to account for the realistic phase-space correlations.

The evolution of the longitudinal phase space also provides a useful diagnostic for the optimization. 
When \(V_{1,0}\) is smaller than the optimal value, the longitudinal phase space exhibits a banana-like shape, with higher-energy particles located toward the head of the bunch. 
At the optimal voltage, the positive- and negative-energy portions of the distribution become approximately symmetric, indicating that the dominant linear correlation has been properly balanced. 
If \(V_{1,0}\) is increased further, the correlation is over-corrected, and the lower-energy particles move toward the head of the bunch. 
This behavior indicates that optimal compression is achieved near the transition between under-correction and over-correction, where the longitudinal phase space is closest to being symmetric.

After the optimal value \(V_{1,0,\mathrm{opt}}\) is obtained, we further scan the phase of the C-band cavity. 
In this step, the K-band linearizer is turned on. 
For each value of the C-band cavity phase \(\phi_1\), the voltage of the C-band cavity and that of the K-band linearizer are constrained by Eqs.~\ref{eq:rf_energy_offset} and \ref{eq:V1_from_h}: 
\begin{equation}
V_1 =
\frac{V_{1,0,\mathrm{opt}}}{\cos\phi_1},
\label{eq:V1_phase_scan}
\end{equation}
and
\begin{equation}
V_2 =
- V_1 \sin\phi_1 .
\label{eq:V2_phase_scan}
\end{equation}
With this choice, the linear energy chirp from the C-band cavity is kept approximately fixed at its optimized value, while the K-band linearizer removes the net energy offset of the reference particle and provides a controllable second-order chirp. 
The scan of \(\phi_1\) therefore mainly optimizes the nonlinear compensation without substantially changing the linear compression condition.

\begin{figure}
 \centering
        \includegraphics[width=0.82\textwidth]{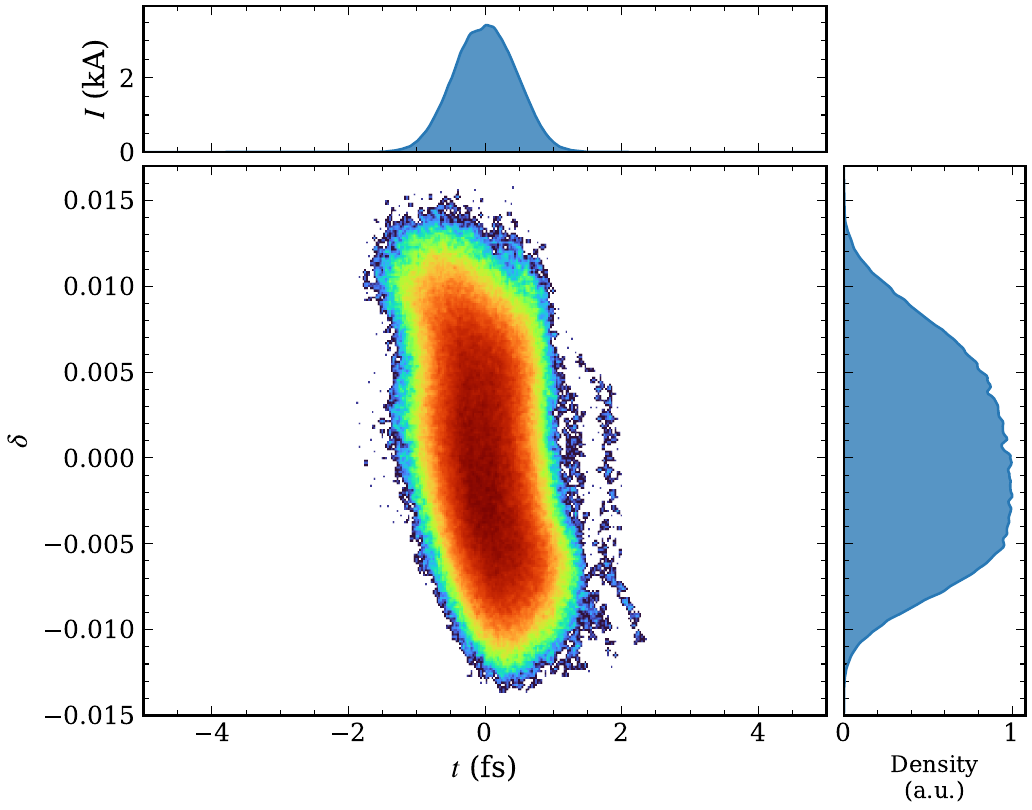}
 \caption{
Final longitudinal properties of the compressed start-to-end beam at the
optimized RF setting. The central panel shows the longitudinal phase-space
density in \((t,\delta)\). The upper projection is the current profile, with a
peak current of approximately \(3.5~\mathrm{kA}\), and the right projection is
the normalized relative-momentum distribution.
}
\label{fig7}
\end{figure}

At the optimized RF setting, the proposed compressor achieves a compression factor of approximately \(400\). The phase scan in Fig.~\ref{fig6} gives a minimum bunch length of \(0.45~\mathrm{fs}\). The corresponding current and energy profiles are presented in Fig.~\ref{fig7}, showing a peak current of approximately \(3.5~\mathrm{kA}\).

To evaluate the robustness of the proposed compression scheme, we further study the influence of beamline-parameter jitter. 
To represent realistic stability errors, random jitters are applied to the RF system, magnet settings, beam energy, and bunch charge. 
For each jittered setting, the beam is tracked through the full compression beamline, and the final bunch length is evaluated.

\begin{figure}
 \centering
        \includegraphics[width=0.98\textwidth]{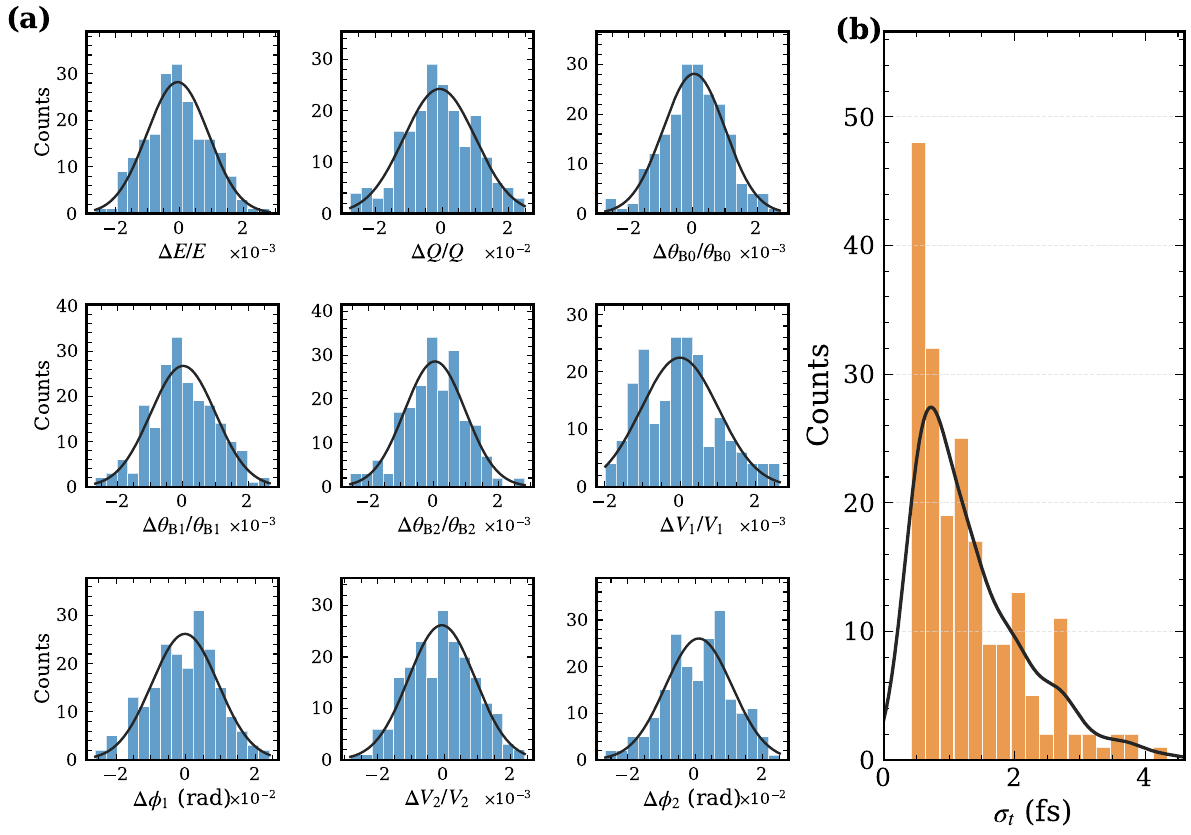}
 \caption{
Jitter tolerance study at \(200~\mathrm{MeV}\) using 200 random error seeds.
(a) Sampled input errors in the beam momentum and charge, the three dipole
bending angles, and the voltages and phases of the C-band and K-band cavities.
The blue histograms show the sampled errors, and the black curves are Gaussian
fits. (b) Distribution of the final bunch length after full-beamline
tracking. The orange histogram shows the simulation results, and the black
curve is a kernel-density estimate.
}
\label{fig8}
\end{figure}

{The results are shown in Fig.~\ref{fig8}. For the 200 random
error seeds, the output distribution remains centered close to the
sub-femtosecond regime, while bunch lengths longer than \(4~\mathrm{fs}\) are
rare. These results indicate that the proposed two-dimensional compression
scheme retains a reasonable tolerance to the combined RF, magnetic, beam-energy,
and bunch-charge fluctuations considered here.}

{
To further examine the energy dependence using start-to-end beam distributions,
we keep the electron-gun and low-energy settings unchanged and adjust the
total length of the downstream travelling-wave accelerating modules. ASTRA
simulations are then used to generate \(4~\mathrm{pC}\) beams at nominal
energies of \(100\), \(200\), \(350\), \(500\), \(750\), and
\(1000~\mathrm{MeV}\). The resulting non-Gaussian phase-space distributions
are matched to the compressor. At each energy, the C-band and K-band RF parameters are
re-optimized following the two-step procedure described above.
\par
}

\begin{figure}[t]
\centering
\includegraphics[width=0.94\textwidth]{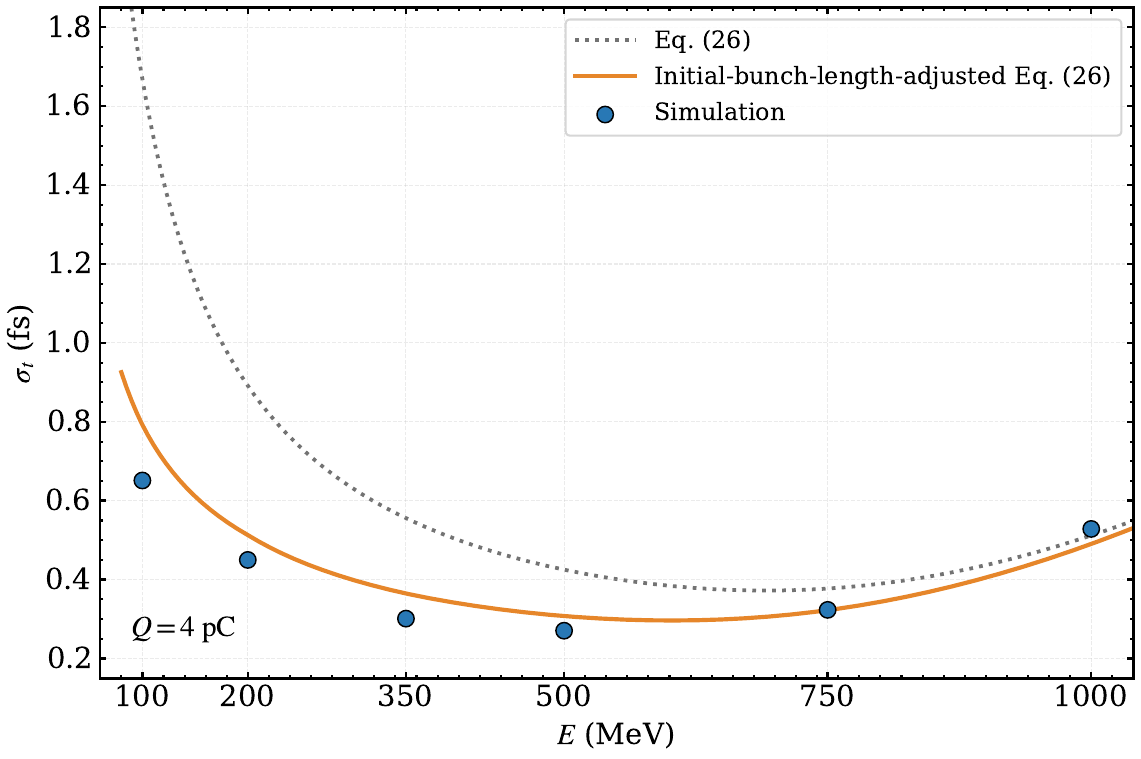}
\caption{{
Comparison of optimized \(4~\mathrm{pC}\) start-to-end beam simulations with
the complete bunch-length theory of Eq.~\ref{eq:complete_bunch_length}. The grey dotted curve directly
uses Eq.~\ref{eq:complete_bunch_length} with the original Gaussian-beam coefficient
\(k_{\mathrm{CSR}}^{\mathrm{G}}\), while the orange curve uses
Eq.~\ref{eq:complete_bunch_length} with
the initial-bunch-length correction in
Eq.~\ref{eq:kcsr_s2e_correction}. The blue symbols show the optimized
simulation results.
}}
\label{fig:s2e_energy_comparison}
\end{figure}

{
The optimized bunch lengths are approximately \(0.65\), \(0.45\),
\(0.30\), \(0.27\), \(0.32\), and \(0.53~\mathrm{fs}\) at the six respective
energies, as shown in Fig.~\ref{fig:s2e_energy_comparison}. The nonmonotonic
trend is characterized by a decrease in bunch length up to approximately
\(500~\mathrm{MeV}\), followed by an increase toward \(1~\mathrm{GeV}\).
This behavior reflects the competition between the ideal and CSR
contributions, both of which decrease with increasing energy, and the ISR
contribution, which grows rapidly and becomes important at high energy. The
observed trend is therefore consistent with the model developed in
Sec.~\ref{sec:length_charge_energy}.
\par
}

{
The initial bunch lengths of the start-to-end beams are approximately
\(138\)--\(144~\mathrm{\mu m}\), substantially longer than the
\(60~\mathrm{\mu m}\) Gaussian bunch used to determine
\(k_{\mathrm{CSR}}\) in Sec.~\ref{sec:length_charge_energy}. Because the CSR
kick scales approximately as \(\sigma_z^{-4/3}\), we correct the coefficient
for each start-to-end beam according to
\begin{equation}
k_{\mathrm{CSR}}^{\mathrm{S2E}}
=
k_{\mathrm{CSR}}^{\mathrm{G}}
\left(
\frac{\sigma_{z,\mathrm{G}}}
{\sigma_{z,\mathrm{S2E}}}
\right)^{4/3},
\label{eq:kcsr_s2e_correction}
\end{equation}
where the superscripts \(\mathrm{G}\) and \(\mathrm{S2E}\) denote the
Gaussian and start-to-end beams, respectively. Without this correction, the
Gaussian-beam theory systematically overestimates the bunch length at low and
intermediate energies. After the initial-length correction, the theoretical
curve reproduces all six optimized simulations with a maximum discrepancy of
approximately \(18\%\). This agreement confirms that the energy dependence
predicted in Sec.~\ref{sec:length_charge_energy} remains applicable to
non-Gaussian start-to-end beams when their different initial bunch lengths are
included.
\par
}

\clearpage

{
Finally, we repeat the jitter study for the start-to-end beams at all six energies.
The optimized solution at each energy is used as the nominal working point,
and the same set of 200 random error vectors and the same jitter definitions
as in the \(200~\mathrm{MeV}\) case are applied to the RF system, dipole
strengths, incoming beam energy, and bunch charge. This procedure isolates the
energy dependence of the statistical response without changing the assumed
technical stability levels.
\par
}

\begin{figure}[htbp]
\centering
\includegraphics[width=0.94\textwidth]{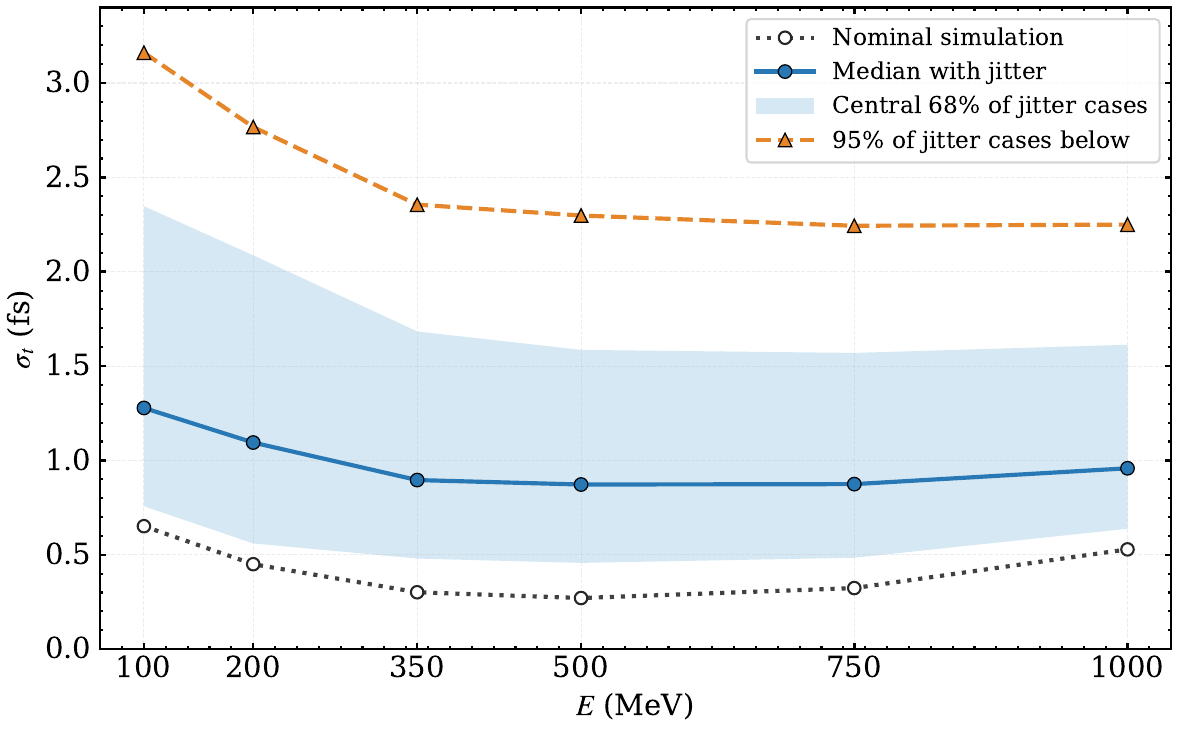}
\caption{{
Energy dependence of the jittered output bunch length for the
\(4~\mathrm{pC}\) start-to-end beams. At each energy, 200 random error seeds
are tracked using the same jitter definitions. The black dotted curve denotes
the nominal optimized simulation with open circles, the blue solid curve with
circles gives the median of the jittered distribution, and the blue shaded
band contains the central (68\%) of the jittered results. The orange dashed
curve with triangles marks the level below which (95\%) of the jittered
results fall.
}}
\label{fig:jitter_energy_dependence}
\end{figure}

{
As shown in Fig.~\ref{fig:jitter_energy_dependence}, the median jittered bunch
length decreases from approximately \(1.28~\mathrm{fs}\) at
\(100~\mathrm{MeV}\) to about \(0.88~\mathrm{fs}\) at
\(350\)--\(750~\mathrm{MeV}\), before increasing slightly to
\(0.96~\mathrm{fs}\) at \(1~\mathrm{GeV}\). The bunch-length range
containing the middle 68\% of the jitter cases contracts markedly between
\(100\) and \(350~\mathrm{MeV}\) and then remains nearly unchanged up to
\(1~\mathrm{GeV}\). Similarly, the bunch length below
which 95\% of the jitter cases fall decreases from approximately
\(3.16~\mathrm{fs}\) at \(100~\mathrm{MeV}\)
to \(2.35~\mathrm{fs}\) at \(350~\mathrm{MeV}\), and subsequently approaches
a plateau near \(2.25~\mathrm{fs}\). Although the nominal bunch length
increases at \(1~\mathrm{GeV}\), consistent with the growing ISR contribution
discussed in Sec.~\ref{sec:length_charge_energy}, the jitter statistics change
only modestly above several hundred MeV. Thus, the largest improvement in
jitter response occurs from low to intermediate energy, after which the
statistical sensitivity to beam energy is substantially reduced.
\par
}

\clearpage

\section{Conclusion}
\label{sec:conclusion}

In this work, we have proposed a two-dimensional beam-compression scheme for generating sub-femtosecond electron bunches at hundred-MeV energies and pC-level charge. The method is based on transverse--longitudinal coupling, in which dispersive beam optics convert the small transverse emittance of modern electron beams into an ultrashort bunch length. Compared with conventional longitudinal compression, the proposed scheme can reach the sub-femtosecond regime with a much smaller correlated energy chirp, thereby reducing the RF-voltage requirement and making the approach more suitable for high-energy beams.

A linear transfer-map analysis was developed to clarify the compression mechanism. We showed that, after the dominant longitudinal and energy-spread contributions are cancelled, the final bunch length is governed by the transverse beam quality. This result highlights the central advantage of the method: the compressed bunch length is linked to the geometric emittance rather than directly to the initial energy spread. We also compared different dispersive-section geometries and found that the three-dipole section provides a larger feasible parameter space and a stronger compression capability than a conventional dogleg. For large bending angles, nonlinear transport effects become important, and a K-band linearizer was introduced to compensate the dominant second-order contribution. Zero-current simulations confirmed that this compensation restores the compressed bunch length close to the ideal linear limit.

{
The effects of CSR, ISR, and space charge were further quantified over a broad range of bunch charges and beam energies. Comparisons of one- and two-dimensional CSR models and of longitudinal and three-dimensional space-charge models show that the reduced models accurately reproduce the final bunch length, with CSR dominating the charge-dependent degradation in the low- to intermediate-energy range considered. We then developed a complete scaling model in which the ideal bunch length decreases as \(\gamma^{-1/2}\), the effective CSR contribution scales as \(Q/\gamma\), and the ISR contribution grows as \(\gamma^{5/2}\). The model agrees closely with the simulations and reveals a transition from CSR-dominated to ISR-dominated operation as the beam energy increases. The resulting \(Q\)--\(E\) operating map provides a practical estimate of the sub-femtosecond regime; for the present compressor configuration, the theoretical \(1~\mathrm{fs}\) boundary reaches a maximum charge of approximately \(22.9~\mathrm{pC}\) near \(1.07~\mathrm{GeV}\), while ISR sets an upper energy of approximately \(1.37~\mathrm{GeV}\) in the limit of vanishing charge.
\par
}

Start-to-end simulations were performed for a realistic photocathode-injector-to-compressor beamline. After optimization of the RF parameters, a 200 MeV, pC-level electron bunch was compressed to a bunch length of 0.45 fs, with a peak current of about 3.5 kA. {Jitter studies over beam energies from \(100~\mathrm{MeV}\) to \(1~\mathrm{GeV}\) show that the output bunch-length distribution narrows from low to intermediate energies and then varies only weakly, indicating reasonable tolerance to realistic beamline fluctuations over the investigated range.}

These results demonstrate that two-dimensional beam compression provides a feasible route toward compact, high-energy attosecond electron beam sources. The method may support future ultrafast applications requiring simultaneously high beam energy, pC-level charge and sub-femtosecond bunch length, including undulator-based extreme-ultraviolet or x-ray radiation generation, and inverse-Compton-scattering gamma-ray sources. Future work will focus on detailed tolerance optimization, proof-of-principle experiments, and integration of the compressed beam with radiation-generation beamlines.

%
%

\ack{This work was supported by the National Natural Science Foundation of China (Grant No. 12405176).}

\roles{Sample text inserted for demonstration.}

\data{The data cannot be made publicly available upon publication because no suitable repository exists for hosting data in this field of study. The data that support the findings of this study are available upon reasonable request from the authors.}


\end{document}